\documentclass[a4paper,11pt]{article}
\usepackage{pos}

\usepackage[utf8]{inputenc}
\usepackage{bm}
\usepackage{longtable,boldline}
\usepackage{xcolor}
\usepackage{xfrac}
\usepackage{hyperref}
\newcommand{\Nstates}{\ensuremath{N_{\text{st}}}}

\renewcommand{\d}{\text{d}}

\newcommand{\e}[1]{\text{e}^{#1}}
\newcommand{\order}{\mathrm{O}}

\renewcommand{\sfrac}[2]{\ensuremath{#1/#2}}

\newcommand{\CF}{C_{\text{F}}}

\newcommand{\Nf}{N_{\text{f}}}

\newcommand{\MS}{\ensuremath{\overline{\text{MS}}}}
\newcommand{\LQCD}{\Lambda_{\text{QCD}}}

\newcommand{\als}{\alpha_{\text{s}}}

\newcommand{\alsr}{\als(\sfrac{1}{r})}

\newcommand{\ml}{m_{\text{l}}}
\newcommand{\ms}{m_{\text{s}}}
\newcommand{\mc}{m_{\text{c}}}

\newcommand{\mtot}{m_{\text{tot}}}

\newcommand{\HISQlats}{MILC:2010pul, MILC:2012znn, Bazavov:2017lyh}
\newcommand{\str}[1]{} % text slated for deletion (won't show up in PDF file)
\title{Charm mass effects in the static energy\newline computed in 2+1+1 flavor lattice QCD}
\ShortTitle{Charm mass effects in the static energy}

\author*[a]{Johannes Heinrich Weber}
\author[b,c,d]{Nora Brambilla}
\author[e]{Rafael L.~Delgado}
\author[c,f]{Andreas Kronfeld}
\author[g]{Viljami Leino}
\author[h]{Peter Petreczky}
\author[b,i]{Sebastian Steinbei\ss er}
\author[b]{Antonio Vairo}

\affiliation[a]{Institut f\"ur Physik \& IRIS Adlershof, Humboldt-Universit\"at zu Berlin, Zum Gro\ss en Windkanal~6, D-12489 Berlin, Germany}
  
\affiliation[b]{Physik Department, Technische Universit\"at M\"unchen, James-Franck-Stra\ss e~1, D-85748 Garching b.~M\"unchen, Germany}

\affiliation[c]{Institute for Advanced Study, Technische Universit\"at M\"unchen, Lichtenbergstra\ss e~2a, D-85748 Garching b.~M\"unchen, Germany}

\affiliation[d]{Munich Data Science Institute, Technische Universit\"at M\"unchen, Walther-von-Dyck-Stra\ss e~10, D-85748 Garching b.~M\"unchen, Germany}

\affiliation[e]{Universidad Politécnica de Madrid, Nikola Tesla, s/n, 28031-Madrid, Spain}

\affiliation[f]{Particle Theory Department, Theory Division, Fermi National Accelerator Laboratory, Batavia, IL 60510-5011, USA}

\affiliation[g]{Helmholtz Institut Mainz, Johannes Gutenberg-Universit\"at Mainz, D-55099 Mainz, Germany}

\affiliation[h]{Physics Department, Brookhaven National Laboratory, Upton, NY 11973-5000, USA}

\affiliation[i]{Leibniz-Rechenzentrum der Bayerischen Akademie der Wissenschaften, Boltzmannstra\ss e~1, D-85748 Garching b.~M\"unchen, Germany}
\emailAdd{johannes.weber@physik.hu-berlin.de}
\abstract{\vspace*{-1mm}
\textbf{\textsf{TUMQCD Collaboration}}\\[1em]%
We report our analysis for the static energy in (2+1+1)-flavor QCD over a wide range of lattice spacings and several quark masses. 
We obtain results for the static energy out to distances of nearly 1 fm, allowing us to perform a simultaneous determination of the lattice scales $r_2$, $r_1$ and $r_0$ as well as the string tension, $\sigma$. 
While our results for $\sfrac{r_0}{r_1}$ and $r_0$ $\sqrt{\sigma}$ agree with published (2+1)-flavor results, our result for $\sfrac{r_1}{r_2}$ 
differs significantly from the value obtained in the (2+1)-flavor case, likely due to the effect of the charm quark. 
We study in detail the effect of the charm quark on the static energy by comparing our results on the finest lattices with the previously published (2+1)-flavor QCD results at similar lattice spacing. 
The lattice results agree well with the two-loop perturbative expression of the static energy incorporating finite charm mass effects. %
}

\FullConference{%
  The 39th International Symposium on Lattice Field Theory (Lattice2022),\\
  8-13 August, 2022 \\
  Bonn, Germany 
}

\begin{document}
\maketitle

\section{Introduction}
\label{sec:intro}

The energy of a static quark-antiquark pair separated by a distance $r$, $E_{0}(r)$, is a fundamental observable of QCD.
Nonperturbative calculations with lattice gauge theory~\cite{Bali:2000gf} were important in establishing confinement in
QCD and in understanding its interplay with asymptotic freedom.
Confinement manifests itself in the linear rise of $E_{0}(r)$ at large $r$; the corresponding slope is known as the string tension.
At short distances, i.e., when $r \LQCD \ll 1$, it holds that $\alsr \ll 1$ and $E_{0}(r)$ may be expanded as a series
in~$\als$. This perturbative expansion is known up to $\mathrm{N}^3\mathrm{LL}$ level~\cite{Tormo:2013tha} and can be expressed as:
\begin{equation}
    \label{eq:statenergyI}
    \!\!\!E_{0}(r)\! =\! \Lambda\! -\! \frac{\CF \als}{r}\! \left( 1\! +\! \# \als\! +\! \# \als^{2}\! +\! \# \als^{3} \ln\als\! +\! \# \als^{3}\! +\!
        \# \als^{4} \ln^{2}\als\! +\! \# \als^{4} \ln\als\! +\! \dots \right) ,
\end{equation}
where $\Lambda$ is a constant of mass dimension one.
The static energy has been extensively studied in (2+1)-flavor QCD, 
while the study of the static energy in ($2+1+1$)-flavor QCD, is less established~\cite{MILC:2010pul, MILC:2012znn, EuropeanTwistedMass:2014osg}.

The static energy is also an important way to determine the strong coupling, $\als$, or, equivalently, $\Lambda_{\MS}$ by fitting to Eq.~\eqref{eq:statenergyI}
(see Ref.~\cite{Bazavov:2019qoo,Komijani:2020kst,dEnterria:2022hzv} for a review).
Generally, perturbative QCD describes the lattice results well up to distances $r\approx$~0.15--0.2~fm.
However, the inverse charm quark mass, $\sfrac{1}{\mc} = \sfrac{1}{1.28}~\text{GeV}^{-1} \sim 0.15~\text{fm}$, lies within this range, 
so the charm quark can neither be considered massless nor infinitely heavy.
It is important to account for finite charm quark mass effects when analyzing the static energy in $(2+1+1)$-flavor.
Furthermore, the static energy plays also an important role in lattice QCD in setting the lattice scale, 
which is commonly done by calculating the static force $F(r) \equiv \d E_{0}(r)/\d r$ and finding scales such that $r_{i}^{2} F(r_{i})= c_{i}$.
We will consider the scales $r_{i},~i=0,1,2$ with $c_{0}=1.65$~\cite{Sommer:1993ce}, $c_{1}=1$~\cite{Bernard:2000gd}, and $c_{2}=\sfrac{1}{2}$~\cite{Bazavov:2017dsy}.

In this proceedings we summarize our recent paper~\cite{Brambilla:2022het},
where we studied the static energy in ($2+1+1$)-flavor QCD and
determined the scales $\sfrac{r_{0}}{a}$, $\sfrac{r_{1}}{a}$, and $\sfrac{r_{2}}{a}$ simultaneously, and obtained $r_{0}$ and $r_{1}$ in physical units and the ratios $\sfrac{r_{0}}{r_{1}}$ and $\sfrac{r_{1}}{r_{2}}$ in the continuum limit.
We also fit the string tension $\sigma$.
Furthermore, we show the impact of finite charm quark mass effects on the static energy by comparing our new lattice
QCD results for the static energy in ($2+1+1$)-flavor QCD both with perturbation theory and 
with published results in ($2+1$)-flavor QCD~\cite{HotQCD:2014kol,Bazavov:2017dsy} at similar lattice spacings.

\section{Measurement of the scales and string tension}
\label{sec:sim}
To compute the static energy, we employ ($2+1+1$)-flavor lattice ensembles generated by the MILC
Collaboration~\cite{\HISQlats}. 
For gluons the one-loop Symanzik-improved action with tadpole improvement has been used.
The sea quarks, namely two isospin-symmetric light quarks, and physical strange and charm quarks, are simulated with the HISQ-action~\cite{Follana:2006rc}.
Throughout the proceedings, we will denote the ensembles by their respective $\beta$ values and their light quark mass labeled with
roman numerals i, ii, or iii, indicating $\sfrac{\ml}{\ms}$ at the physical value, $\sfrac{1}{10}$, or $\sfrac{1}{5}$,
respectively. 
The gauge configurations have been fixed to Coulomb gauge, which allows for easy access to off-axis distances $r$.
In the characterization of these ensembles, we use the lattice scale $a_{f_{p4s}}$~\cite{MILC:2012znn}. 
For exact technical details about the simulations, we refer the reader to the original paper~\cite{Brambilla:2022het}.

The static energy is obtained from the time dependence of the Wilson-line correlation function
$C\left(\bm{r},\tau,a\right)$ at separation $\sfrac{\bm{r}}{a}$,
that we reparameterize using energy differences $a\Delta_{n}(\bm{r},a)=aE_{n}(\bm{r},a)-aE_{(n-1)}(\bm{r},a)>0$:
\begin{align}
    C\left(\bm{r},\tau,a\right) &= \e{-\tau E_{0}\left(\bm{r},a\right)} \left( C_{0}\left(\bm{r},a\right) +
        \sum\limits_{n=1}^{\Nstates-1} C_{n}\left(\bm{r},a\right)
            \prod\limits_{m=1}^{n} \e{-\tau \Delta_{m}\left(\bm{r},a\right)}
        \right) + \ldots,
    \label{eq:forward_tower}
\end{align}
and choose $\Nstates=1$, $2$, or~$3$ to fit our data to this form using Bayesian priors. 
We use the results of the $\Nstates=1$ fits as the starting guess for the $\Nstates=2$ fits and similarly
for the $\Nstates=3$ fits. The $\Nstates=2$-fit will serve as our main result, with $\Nstates=3$ serving as a check of systematic errors.
For each ensemble, we have also constructed a Wilson-line correlation function replacing the bare gauge links with
links after one iteration of HYP-smearing~\cite{Hasenfratz:2001hp},
which improves the signal-to-noise ratio at large distances.
Again, the reader is referred to the main publication for the technical details~\cite{Brambilla:2022het}.

\begin{figure}
    \centering
    \includegraphics[width=0.485\textwidth,clip,trim=0mm 0mm 5mm 12mm]{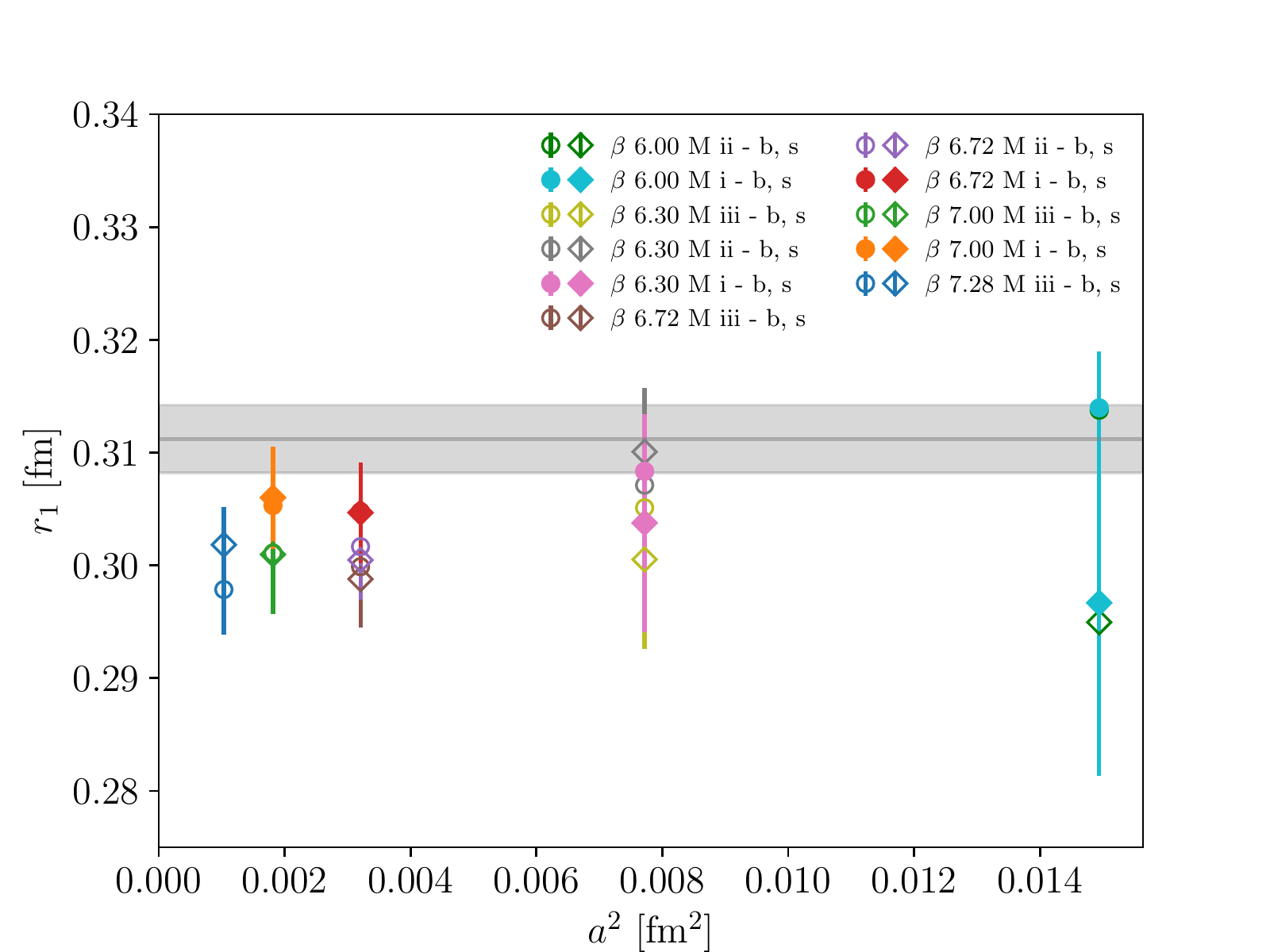}%
    \includegraphics[width=0.475\textwidth,clip,trim=165mm 5mm 0mm 5mm]{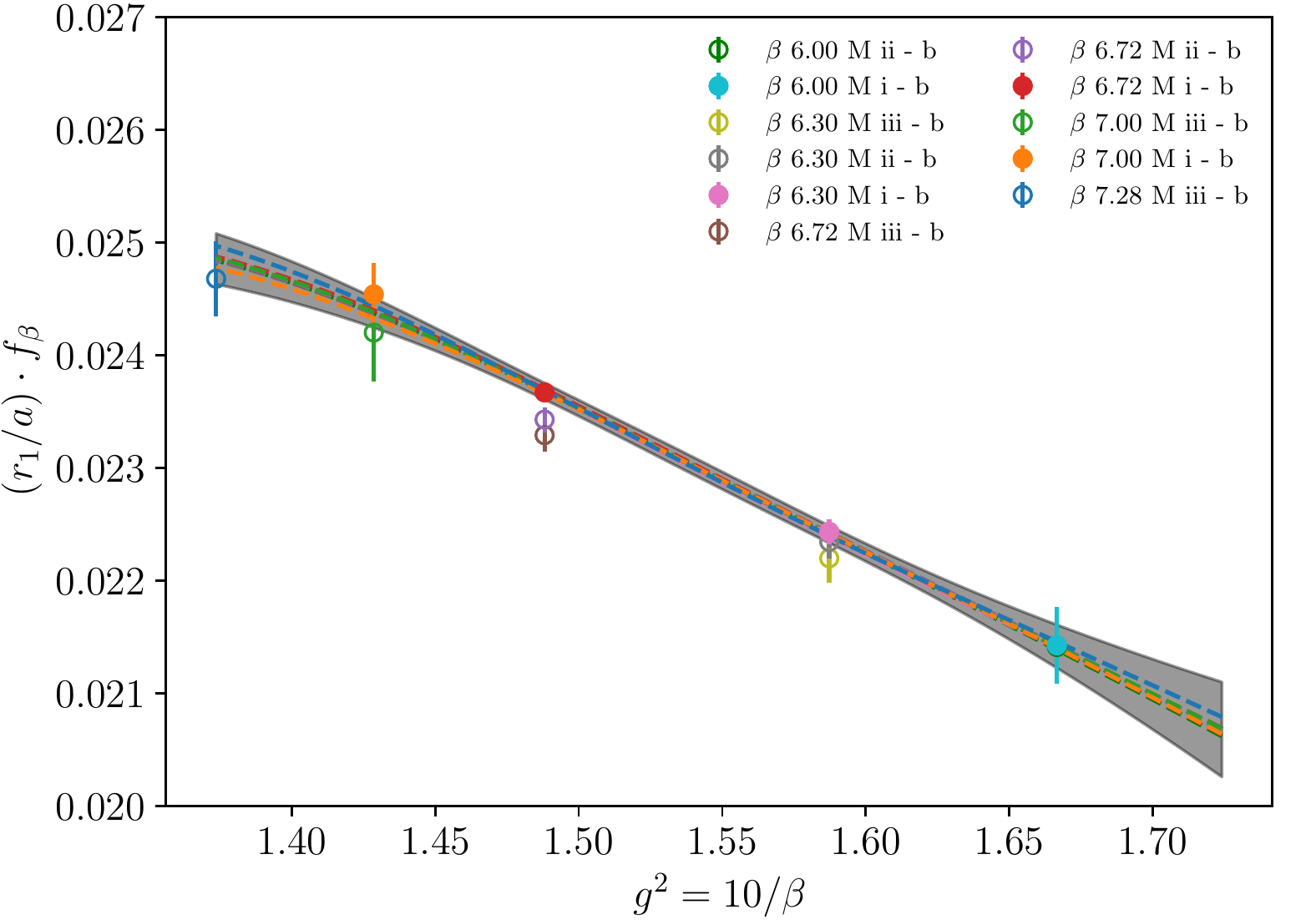}%
    \caption{\label{fig:r_i_over_a_Allton_summary}%
    The scale $r_1$ for all ensembles (colors) and bare ($\circ$) and smeared ($\diamond$) gauge links.
    Left: We use the lattice scale $a_{f_{p4s}}$ to convert to physical units and as an $x$-axis.
    The gray band indicates the ($2+1$)-flavor value from FLAG 2021~\cite{FlavourLatticeAveragingGroupFLAG:2021npn}.
    Right: The scale multiplied by the two-loop $\beta$-function. The curves indicate the Allton fit~\eqref{eq:Allton_fit}, 
    evaluated at the physical mass.
    The color of the lines indicates the ensemble that has been left out, while the black curve includes all ensembles.
    }
\end{figure}%
To determine the lattice scales $r_i$, we must first determine the static force. 
We do this by fitting the Cornell potential $E(R,a) = -\sfrac{A}{R} + B + \Sigma R$ and calculating the derivative analytically.
Here, $R$ is the tree-level improved distance, defined by matching LO formulas of lattice and continuum perturbation theories.
For each scale $r_i$, we perform the Cornell fit locally to randomly picked data points close to the expected physical distance of the respective scale.
This random picking procedure allows us to scope the systematic error raising from directional dependent 
terms not covered by the tree-level improvement.
We repeat the random picking for around hundred times to make sure the picking itself doesn't cause any problems. 
To measure the correlations we use the jackknife pseudoensembles to measure the covariance matrix
and use the same picks for each pseudoensemble. 
As an example, we present our extracted scale $r_1$ in Fig.~\ref{fig:r_i_over_a_Allton_summary}.
Since the correlators with bare- or smeared-link variables represent different discretizations, different values of the
scales $\sfrac{r_{i}}{a}$ with bare or smeared links are to be expected.
This effect needs to be distinguished from the distortions of the smeared-link correlators at small distances due to
the unphysical contact-term interactions.
While the former is not a problem, the latter needs to be avoided. Hence we only use smeared results 
that have data at sufficiently large $R$.

To further smooth the data, we interpolate with an Allton Ansatz~\cite{Allton:1996kr}, that presents the scales
$\sfrac{r_{i}}{a}$ as functions of the squared bare gauge coupling $g_{0}^{2}$ and the bare quark masses
$am_{q}$.
\begin{align}
    \frac{a}{r_{i}} &= \frac{C_{0} f_{\beta} + C_{2} g_{0}^{2} f_{\beta}^{3} + C_{4} g_{0}^{4} f_{\beta}^{3}}
        {1 + D_{2} g_{0}^{2} f_{\beta}^{2}}\,,
    \label{eq:Allton_fit} \\
    C_{0} &= C_{00} + C_{01l} \frac{a\ml}{f_{\beta}} + C_{01s} \frac{a\ms}{f_{\beta}} +
        C_{01} \frac{a\mtot}{f_{\beta}} + C_{02} \frac{(a\mtot)^{2}}{f_{\beta}}, \\
    C_{2} &= C_{20} + C_{21} \frac{a\mtot}{f_{\beta}}, \quad a\mtot = 2 a\ml + a\ms + a\mc\,,
\end{align}
where $f_{\beta}$ is the integrated $\beta$ function to two loops and $C_{ij}$ are fit parameters.
We cannot reliably constrain the coefficients of multiple quark mass dependent terms.
The parametrization yielding smallest reduced $\chi^{2}$ is quadratic in $a\mtot$ with only $C_{00}$, $C_{02}$,
$C_{20}$, and $D_{2}$ being allowed to vary.
Hence,  we use these quadratic fits including all ensembles as our main results. 
Further, we cannot constrain $C_{4}$ and $C_{21}$, so they are set to zero.

In order to test whether the Allton fit might assign an undue, large weight to any ensemble, we repeated the same
Allton fit on each subset of the data leaving out one ensemble in each; all of these fits are covered by the regression
error of the Allton fit using the full data set, see Fig.~\ref{fig:r_i_over_a_Allton_summary}.
The errors of the individual $\sfrac{r_{i}}{a}$ contain our estimates of systematic uncertainties, dominated by the
variation of the independent, randomly chosen sets of $R$ values. These are considerably larger than the statistical errors, hence we add them and the statistical uncertainties in quadrature.
The regression errors of the smoothened $\sfrac{r_{i}}{a}$ therefore reflect the systematic errors.

Next we perform the continuum extrapolations for all quantities.
Knowing that the leading discretization effects are of order $\als^{2}a^{2}$ and $a^{4}$,
we will formulate a set of functional forms for the continuum extrapolation:
\begin{align*}
     \xi_{0}\! &  &&\text{(avg)} , 
    &\xi_{0}\! &+\! \alpha^{2}  \xi_{1} x && \text{(lin)} , \\
     \xi_{0}\! &+\! \alpha^{2}  \xi_{1} x\! +\! \xi_{2} x^{2} && \text{(quad)} , 
    &\xi_{0}\! &+\! \alpha^{2} [\xi_{1} x\! +\! \xi_{2} x y] && \text{(l,lm)} , \\
     \xi_{0}\! &+\! \alpha^{2} [\xi_{1} x\! +\! \xi_{2} x y]\! +\! \xi_{3} x^{2} && \text{(q,lm)} , 
    &\xi_{0}\! &+\! \alpha^{2} [\xi_{1} x\! +\! \xi_{2} x y^{2}] && \text{(l,qm)} , \\
     \xi_{0}\! &+\! \alpha^{2} [\xi_{1} x\! +\! \xi_{2} x y^{2}]\! +\! \xi_{3} x^{2} && \text{(q,qm)} , 
    &\xi_{0}\! &+\! \alpha^{2} [\xi_{1} x\! +\! \xi_{2} x y]\! +\! \xi_{3} y && \text{(l,lm,mc)} , \\
     \xi_{0}\! &+\! \alpha^{2} [\xi_{1} x\! +\! \xi_{2} x y]\! +\! \xi_{3} x^{2}\! +\! \xi_{4} y && \text{(q,lm,mc)} , 
    &\xi_{0}\! &+\! \alpha^{2} [\xi_{1} x\! +\! \xi_{2} x y^{2}]\! +\! \xi_{3} y && \text{(l,qm,mc)} , \\
     \xi_{0}\! &+\! \alpha^{2} [\xi_{1} x\! +\! \xi_{2} x y^{2}]\! +\! \xi_{3} x^{2}\! +\! \xi_{4} y && \text{(q,qm,mc)} ,
\end{align*}
where $x=(\sfrac{a}{r_{0}})^{2}$ or $(\sfrac{a}{r_{1}})^{2}$ represents the lattice spacing dependence and
$y=\sfrac{(a\ml)_{\text{sea}}}{(a\ms)_{\text{sea}}}$ %or $\sfrac{(a\ml)_{\text{sea}}}{(a\ms)_{\text{tuned}}}$ 
represents the light quark mass dependence. 
Further, we assume either $\alpha=1$ or include the tadpole factors  $\alpha = \sfrac{g_{0}^{2}}{(4\pi u_{0}^{4})}$.
\begin{figure}
    \centering
    \includegraphics[width=0.49\textwidth,clip, trim=5mm 5mm 160mm 5mm]{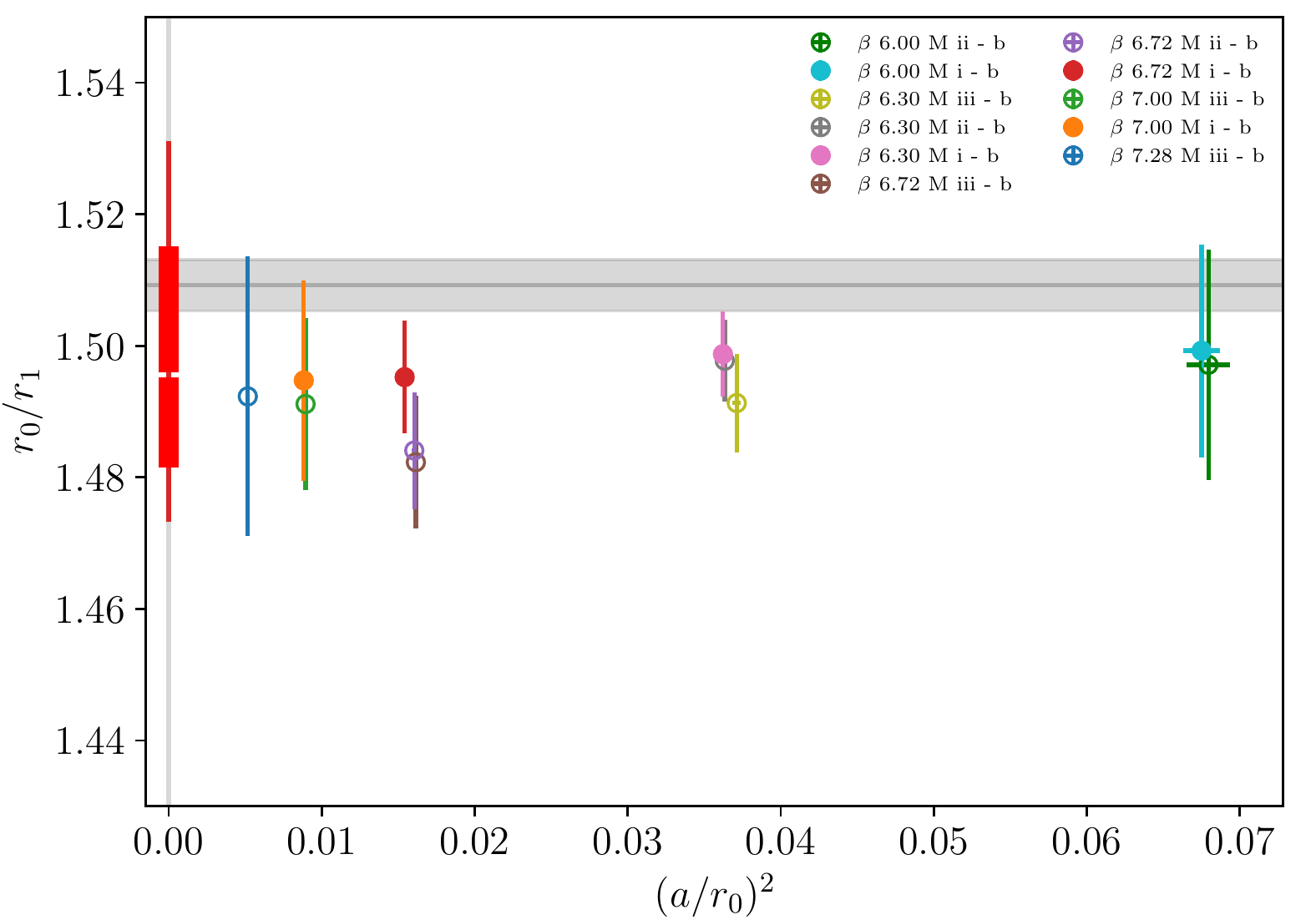}
    \hfill
    \includegraphics[width=0.49\textwidth,clip, trim=5mm 5mm 160mm 5mm]{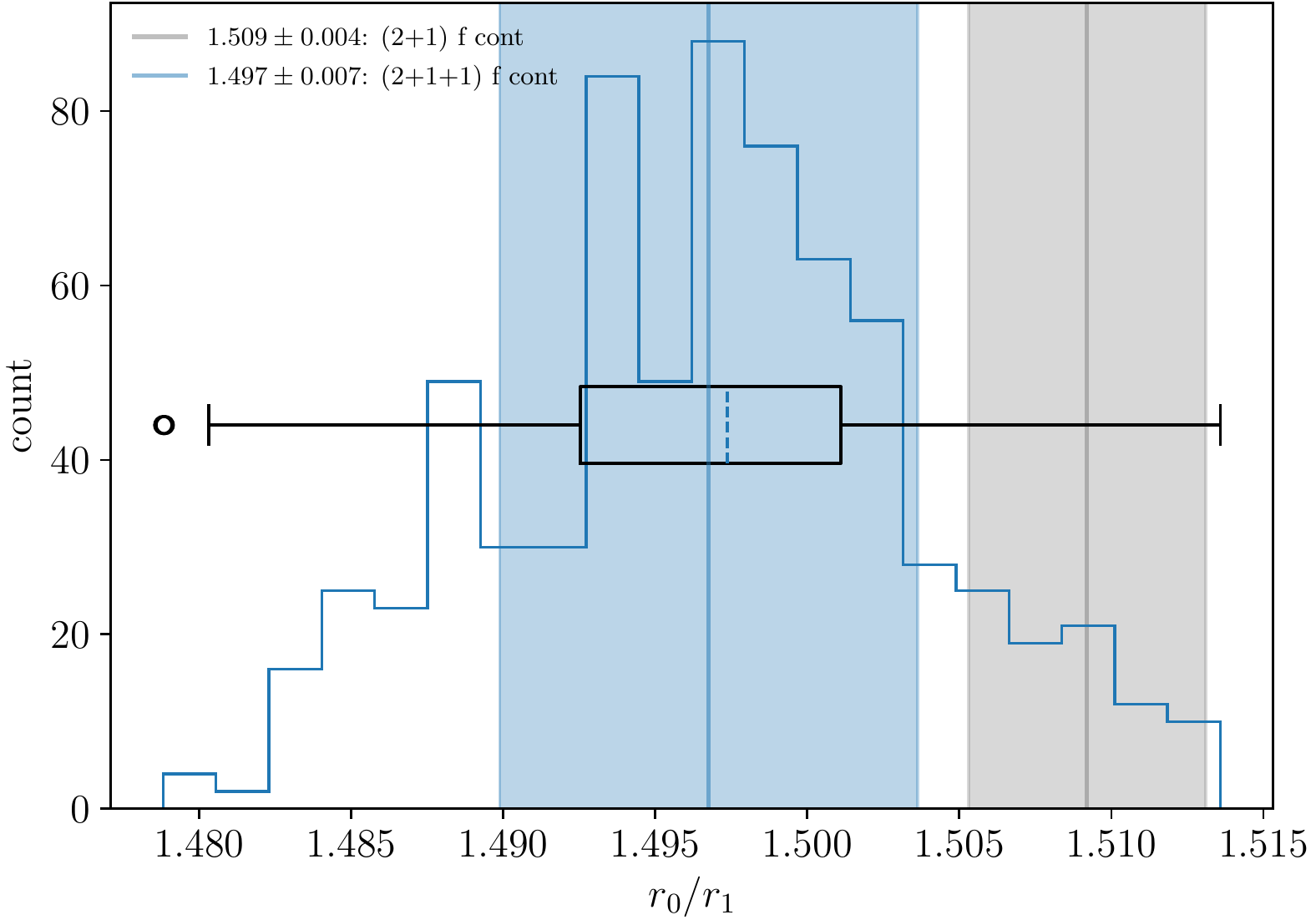}%
    \caption{\label{fig:continuum_ri_rj_III}%
    Left: Continuum results ($\Box$) and smoothened data ($\circ$) for the ratio $\sfrac{r_{0}}{r_{1}}$  using bare links.
    Right: Histogram of the continuum extrapolations for the ratios $\sfrac{r_{0}}{r_{1}}$ from the left side.
    The gray bands indicate the ($2+1$)-flavor QCD value~\cite{HotQCD:2014kol}.
    }
\end{figure}%
For further details about the fits, we refer the reader to the main paper~\cite{Brambilla:2022het}.
Overall, we end up with around hundred fits that are shown in Fig.~\ref{fig:continuum_ri_rj_III} for the ratio of scales $\sfrac{r_{0}}{r_{1}}$.
We then take a weighted average of the histogram as our final continuum result.

\begin{figure}
    \centering
    \begin{minipage}[t]{0.49\textwidth}\vspace{0pt}%
    \includegraphics[width=1.0\textwidth,clip, trim=5mm 5mm 5mm 5mm]{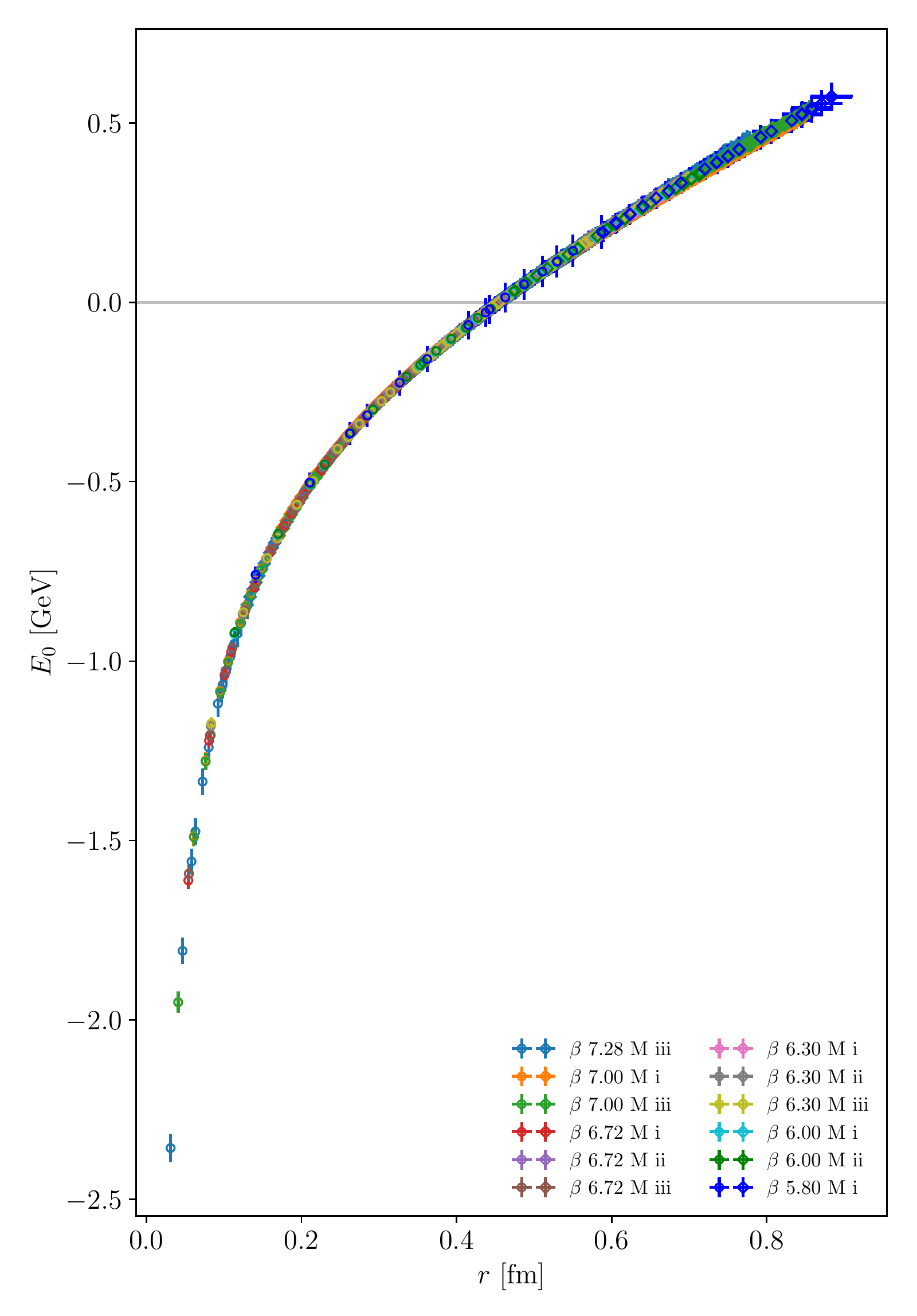}
    \end{minipage}
    \begin{minipage}[t]{0.49\textwidth}\vspace{-0.6em}%
    \centering
    \includegraphics[width=0.65\textwidth,clip, trim=5mm 5mm 5mm 5mm]{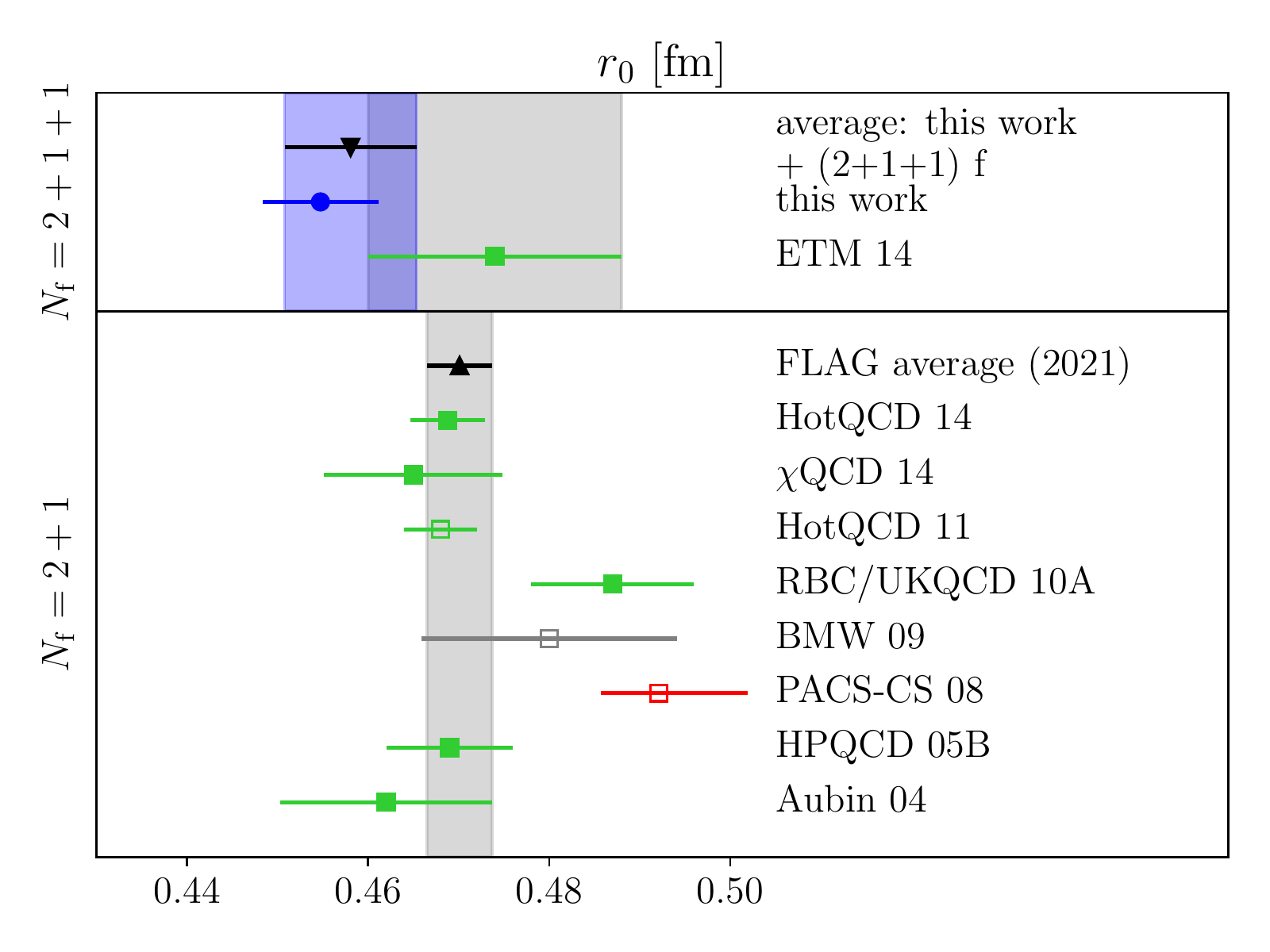} \\
    \includegraphics[width=0.65\textwidth,clip, trim=5mm 5mm 5mm 5mm]{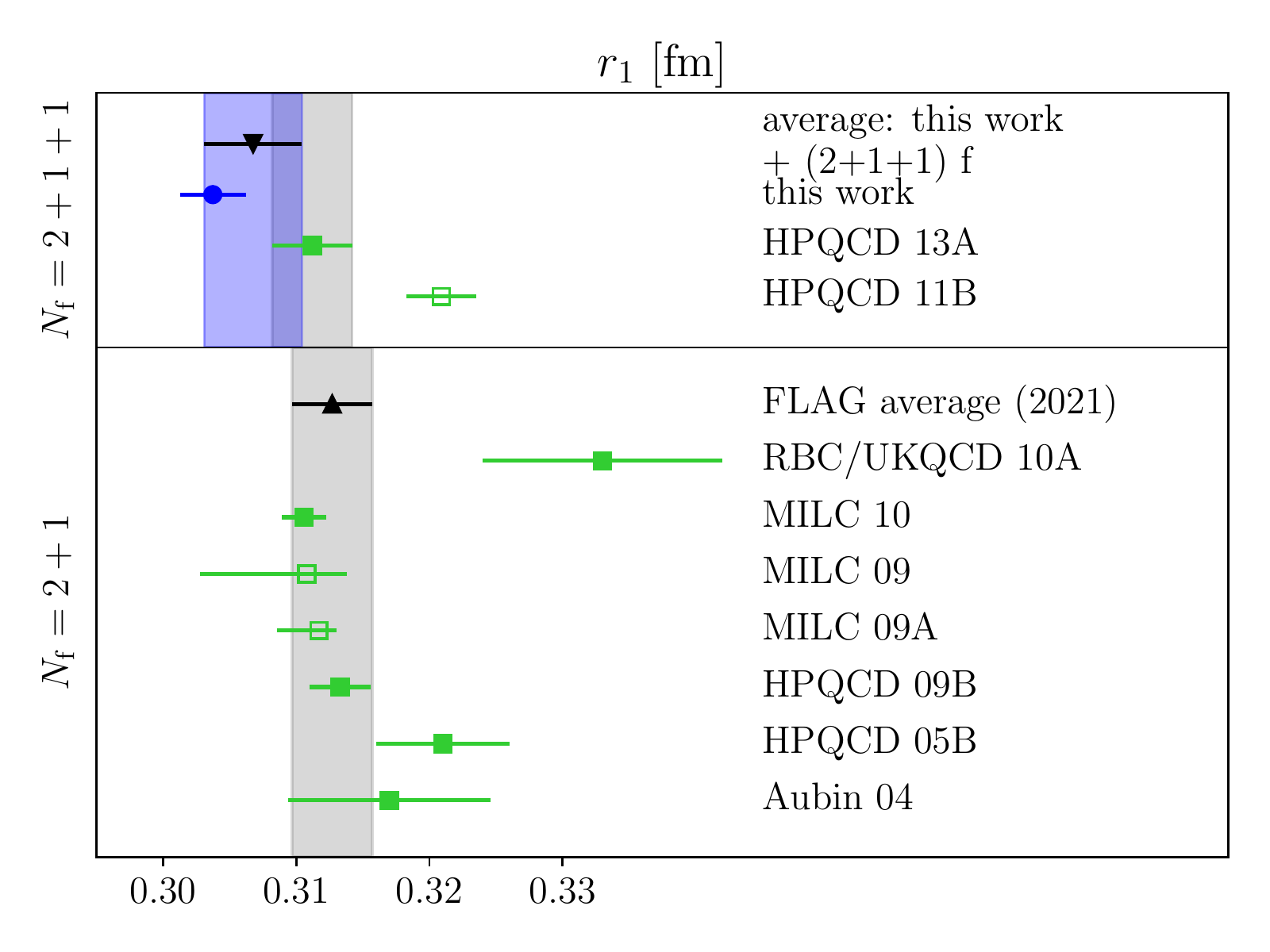} \\
    \includegraphics[width=0.65\textwidth,clip, trim=5mm 5mm 5mm 5mm]{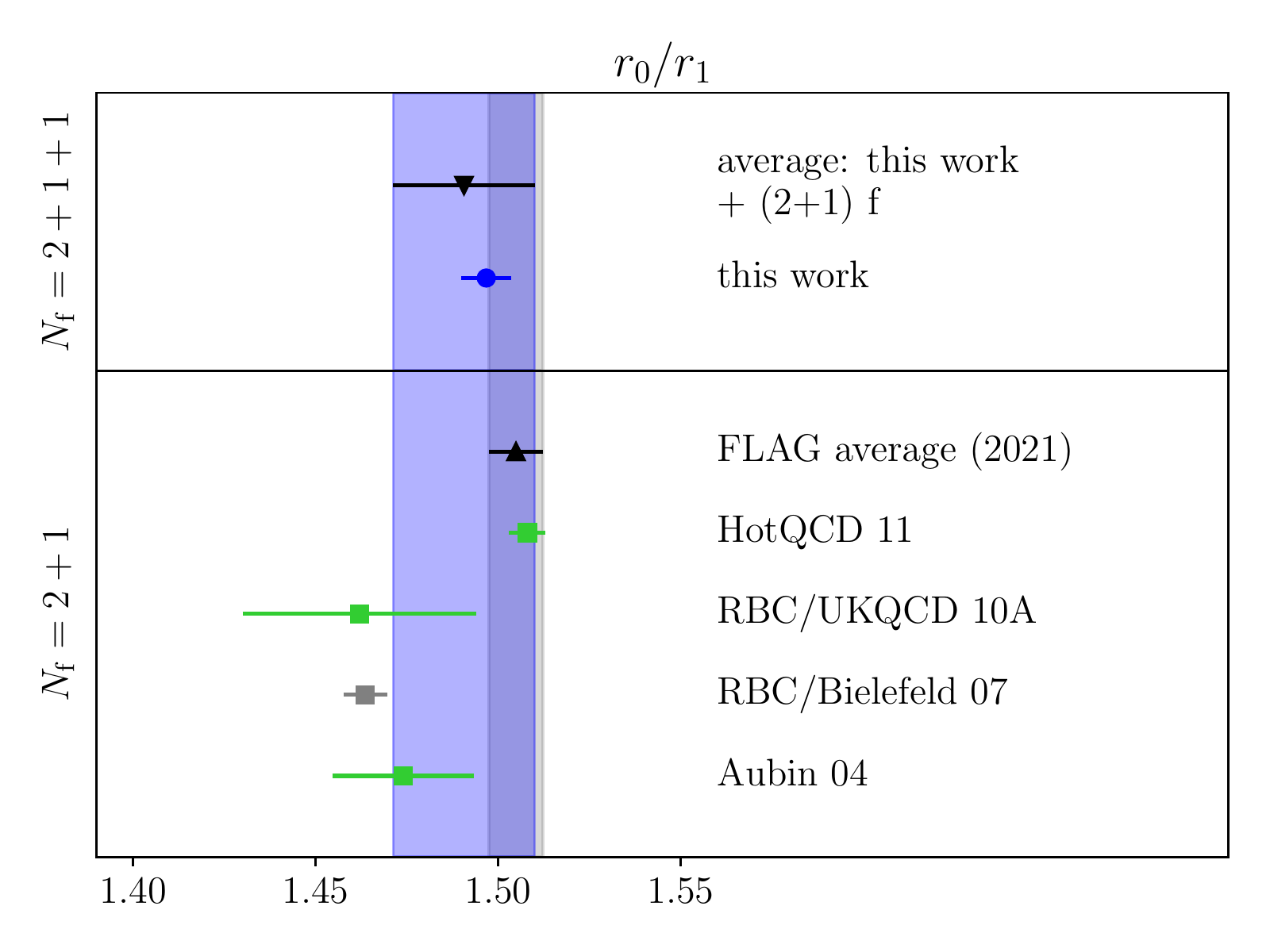}  
    \end{minipage}
    \caption{%
    Left: Our results for the static energy in physical units.
    Right: Comparison plots for $\sfrac{r_{0}}{r_{1}}$, $r_{0}$, and $r_{1}$ 
    with the FLAG 2021 averages~\cite{FlavourLatticeAveragingGroupFLAG:2021npn}.
    The blue bands indicate how the averages would change if our new data is included.
    }
    \label{fig:curve_collapse}
\end{figure}

The steps to final continuum results are similar for all the quantities. 
We calculate the continuum limit for the scales $r_0$ and $r_1$ and their ratios $\sfrac{r_0}{r_1}$ and $\sfrac{r_1}{r_2}$.
The resulting histograms are roughly Gaussian. $\sfrac{r_1}{r_2}$ is an exception whose slight bimodality surfaces possibly due to discretization effects affecting the $\sfrac{r_2}{a}$ extraction on the respective coarsest lattice.
For the ratios we get as a final continuum result
$\sfrac{r_{0}}{r_{1}} = 1.4968 \pm 0.0069$ and $\sfrac{r_{1}}{r_{2}} = 2.313 \pm 0.069$.
Further, we compare $\sfrac{r_{0}}{r_{1}}$ to existing (2+1) results on the right side of Fig.~\ref{fig:curve_collapse}.
The result for $\sfrac{r_1}{r_2}$ is somewhat larger than the (2+1) equivalent~\cite{Bazavov:2017dsy}, which apart from discretization effects in $r_2$, 
is a sign of a charm quark effect.
For the scales themselves, we get $r_{0} = 0.4547 \pm 0.0064~\text{fm}$, $r_{1} = 0.3037 \pm 0.0025~\text{fm}$, 
and $r_{2} = 0.1313 \pm 0.0041~\text{fm}$ (reconstructed from $r_{1}$ and $\sfrac{r_{1}}{r_{2}}$). Again, we show the comparison to existing results both in (2+1)- and (2+1+1)-flavor theories
in Fig.~\ref{fig:curve_collapse}, for $r_0$ and $r_1$.

With the continuum scales set we can demonstrate how the static energy progresses from the Coulombic region to confining region.
Our full extracted static energy, is shown on left side of Fig.~\ref{fig:curve_collapse}.
We show only the  bare (smeared) data for $\sfrac{r}{a} \le 4$ ($\sfrac{r}{a} > 4$).
On the scale of Fig.~\ref{fig:curve_collapse}, it is possible to see light quark mass dependence only at the larger $R$
but it is very difficult to spot lattice-spacing dependence.

Turning to the string tension, our data are insufficient to constrain the coefficient $A$ in the Cornell potential 
when fitting the static energy over the range $r\ge0.58$~fm. This range lies between the Coulomb and (asymptotic) string regime, where a $\sfrac{1}{R}$ behavior is also expected~\cite{Luscher:1980ac}. 
With no obvious physical origin for a $\sfrac{1}{R}$ term in this range, we choose fits fixing $A$ to either
$A_{r_{0}}$, the fit results from the $r_{0}$ fit, or $\sfrac{\pi}{12}$~\cite{Luscher:1980ac}.
In fact, $A_{r_{0}}$ turns out to be within a factor of two of $\sfrac{\pi}{12}$.
We get, for the final continuum limit: $\sqrt{\sigma r_{0}^{2}} = 1.077 \pm 0.016$ for $A = A_{r_{0}}$ and
$\sqrt{\sigma r_{0}^{2}} = 1.110 \pm 0.016$ for $A = \sfrac{\pi}{12}$.

\section{Charm effects}
\label{sec:charm}
Now that we have data for the static energy in ($2+1+1$)-flavor QCD, it is possible to study the effect of the massive charm loops.
Effects due to the finite mass of a heavy quark, while keeping $\Nf$ quarks massless, can be cast into a correction
$\delta V_{m}^{(\Nf)}(r)$ to be added to the static energy.
This correction has been computed at $\order(\als^{3})$; for summary of perturbative results, see Ref.~\cite{Recksiegel:2001xq}.
In our case of interest, the relevant massive quark is the charm quark $\mc = 1.28$~GeV.
The full expression for the static energy in perturbation theory then becomes:
\begin{equation}
    \label{eq:full_statenergy}
    E^{(\Nf)}_{0,m}(r) = \int\limits_{r^{\ast}}^{r} \d r' \; F^{(\Nf)}(r') + \delta V_{m}^{(\Nf)}(r) + \text{const},
\end{equation}
where $\Nf=3$ is the number of massless quarks. 
In the limit $m \gg \sfrac{1}{r}$, Eq.~\eqref{eq:full_statenergy} reduces to theory with $\Nf$ massless quarks $E^{(\Nf)}_{0}(r)$,
while in the limit $m \ll \sfrac{1}{r}$ it reduces to theory with $\Nf+1$ massless quarks $E^{(\Nf+1)}_{0}(r)$.
This is a consequence of the decoupling of the static potential.

\begin{figure}
    \centering
    \includegraphics[width=0.46\textwidth,clip,trim=5mm 4mm 5mm 5mm]{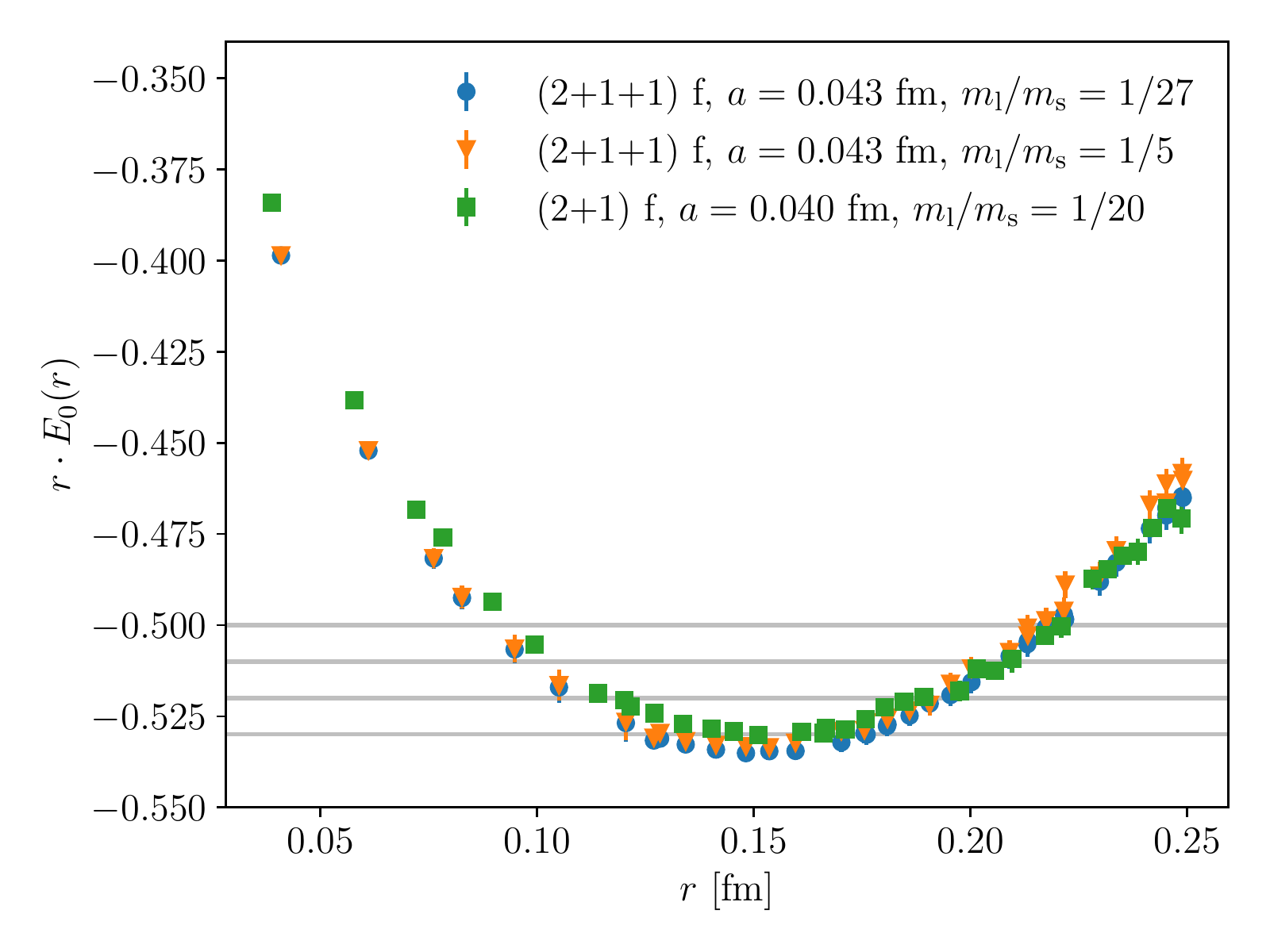}%
    \hfill%
    \includegraphics[width=0.495\textwidth,clip,trim=5mm 5mm 5mm 5mm]{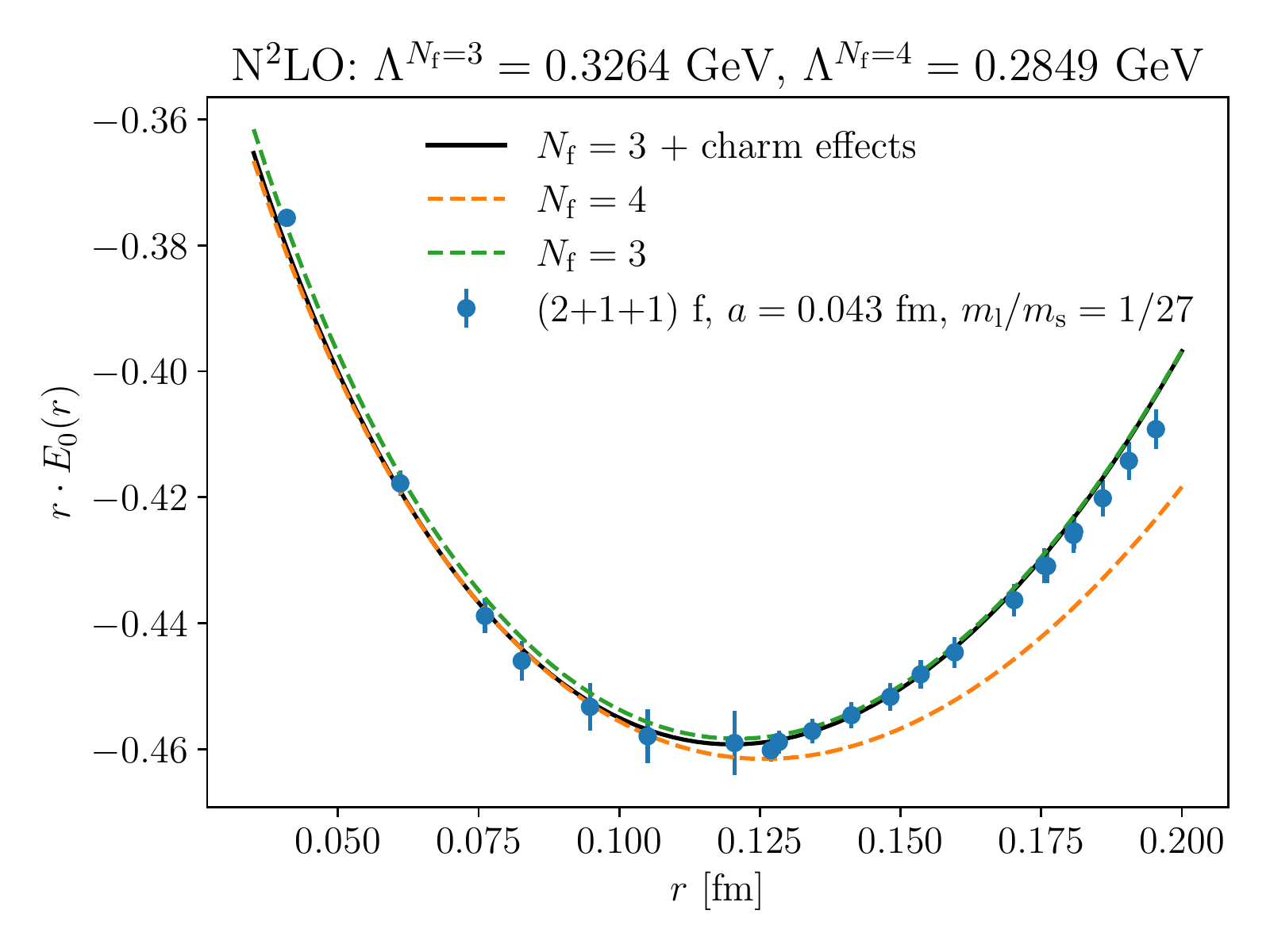}%
    \caption{\label{fig:charm2}%
    Left: The dimensionless quantity $rE_{0}(r)$ for two different ($2+1+1$)-flavor ensembles using different light quark
        masses and one ($2+1$)-flavor ensemble of similar lattice spacing.
    Right: Comparison of the ($2+1+1$)-flavor data with curves obtained from different perturbative expressions of the static
    energy times the distance.
    }
\end{figure}
On the left side of Fig.~\ref{fig:charm2}, we show ($2+1$)-flavor~\cite{HotQCD:2014kol,Bazavov:2017dsy} and ($2+1+1$)-flavor lattice data for $rE_{0}(r)$,
which correspond to different discretizations and to light quark mass over strange quark mass ratios $\sfrac{\ml}{\ms} = \sfrac{1}{20}$ 
and $\sfrac{\ml}{\ms} = \sfrac{1}{27}$, respectively.
For the ($2+1+1$)-flavor data we use the scale $a_{f_{p4s}}$ to convert to physical units,
while for the ($2+1$)-flavor data, we use the published $r_1$ scale~\cite{HotQCD:2014kol}.
To aid the visualization of the small finite mass effects, we add a mass independent constant to the ($2+1+1$)-flavor $E_{0}(r)$ such that the shifted data at physical mass 
are rather flat in the range of interest.
We additionally show another ($2+1+1$)-flavor data set with larger light quark mass,
$\sfrac{\ml}{\ms} = \sfrac{1}{5}$, whose data set has not been shifted relative to the physical one.
We match the ($2+1$)-flavor data to the ($2+1+1$)-flavor data of the similar $\sfrac{\ml}{\ms}$-ratio,
at large distances, $r \gg \sfrac{1}{\mc}
\sim 0.15$~fm, where they must agree up to a constant due to the decoupling of the charm quark.
This matching of the ($2+1$)-flavor data to the ($2+1+1$)-flavor data is done by minimizing their difference over the
range $r \in [0.18,0.27]$~fm and by varying the range to estimate the matching error.
The difference in the light quark mass between the ($2+1$)-flavor data and the ($2+1+1$)-flavor data is smaller than
the one between the two sets of ($2+1+1$)-flavor data.
Since the latter are hardly distinguishable, we deduce that the light quark mass difference should be irrelevant in
this entire range, and that the difference between the ($2+1$)-flavor data and the ($2+1+1$)-flavor data is due to the
dynamical charm quark in the sea.
The effect of the dynamical charm is therefore significant and visible in the data.

In order to compare with perturbation theory, we need to have a value for $\Lambda_{\MS}$.
We determine $\Lambda_{\MS}^{(\Nf=3)}$ by fitting Eq.~\eqref{eq:full_statenergy} using 2-loop expressions everywhere to the physical ($2+1+1$)-flavor ensemble.
We leave out data at $\sfrac{r}{a}=1$ from all the fits and vary the fit range up to $r \approx 0.19$~fm using
$\mc^{\MS}(\mc^{\MS}) = 1.28$~GeV and the three-loop running of $\als$.
The theory curve with massive quarks and curves with $\Nf=3$ and $\Nf=4$ massless quarks, are shown in Fig.~\ref{fig:charm2} in comparison to (2+1+1)-flavor physical ensemble.
We observe that the ($2+1+1$)-flavor lattice data behaves accordingly to the decoupling theorem.
At large distance the data points are well described by the perturbative static energy with $\Nf=3$ massless flavors and at short
distance by the perturbative static energy with $\Nf=4$ massless flavors.
The static energy with three massless flavors and one massive charm interpolates smoothly between these two curves and on the overall describes well the data.

\section*{Acknowledgments}
The simulations were carried out on the computing facilities of the Computational Center for Particle and Astrophysics
(C2PAP) in the project \emph{Calculation of finite $T$ QCD correlators} (pr83pu) and of the SuperMUC cluster at the
Leibniz-Rechenzentrum (LRZ) in the project \emph{The role of the charm-quark for the QCD coupling constant} (pn56bo).
This research was funded by the Deutsche Forschungsgemeinschaft (DFG, German Research Foundation) cluster of excellence
``ORIGINS'' (\href{www.origins-cluster.de}{www.origins-cluster.de}) under Germany's Excellence Strategy
EXC-2094-390783311.
This research is supported by the DFG and the NSFC through funds provided to the Sino-German CRC 110 ``Symmetries and
the Emergence of Structure in QCD''.
R.L.D.\ is supported by the Ramón Areces Foundation, the INFN post-doctoral fellowship AAOODGF-2019-0000329, and the
Spanish grant MICINN: PID2019-108655GB-I00.
Fermilab is managed by Fermi Research Alliance, LLC, under Contract No.\ DE-AC02-07CH11359 with the U.S.\ Department of
Energy.
P.P.\ is supported by the U.S.\ Department of Energy under Contract No.\ DE-SC0012704.
A.V.\ is supported by the EU Horizon 2020 research and innovation programme, STRONG-2020 project, under grant agreement
No.~824093.
J.H.W.’s research is funded by the Deutsche Forschungsgemeinschaft (DFG, German Research
Foundation)---Projektnummer 417533893/GRK2575 ``Rethinking Quantum Field Theory''.
The lattice QCD calculations have been performed using the publicly available
\href{https://web.physics.utah.edu/~detar/milc/milcv7.html}{MILC code}.

\bibliographystyle{JHEP}
\bibliography{ref}

\end{document}